\documentclass[prb,twocolumn,showpacs]{revtex4}
\usepackage{graphicx,epsf}

\newcommand{\bq}{{\bf q}}
\newcommand{\bp}{{\bf p}}
\newcommand{\br}{{\bf r}}
\newcommand{\nubar}{\bar{\nu}}
\newcommand{\rhobar}{\bar{\rho}}
\newcommand{\chibar}{\bar{\chi}}
\newcommand{\Hhat}{\hat{H}}

\begin{document}

\def\tende#1{\,\vtop{\ialign{##\crcr\rightarrowfill\crcr
\noalign{\kern-1pt\nointerlineskip}
\hskip3.pt${\scriptstyle #1}$\hskip3.pt\crcr}}\,}

\title{Microscopic theory of the reentrant integer quantum Hall effect in the first and second excited LLs}
\author{M.\ O.\ Goerbig$^{1,2}$, P.\ Lederer$^2$, and  
C.\ Morais\ Smith$^1$}

\affiliation{$^1$D\'epartement de Physique, Universit\'e de Fribourg, P\'erolles,  CH-1700 Fribourg, Switzerland.\\
$^2$Laboratoire de Physique des Solides, Bat.\,510, Universit\'e Paris-Sud, F-91405 Orsay cedex, France.}

\begin{abstract}
We present a microscopic theory for the recently observed reentrant integer quantum Hall effect in the $n=1$ and $n=2$ Landau levels. Our energy investigations indicate an alternating sequence of $M$-electron-bubble and quantum-liquid ground states in a certain range of the partial filling factor of the $n$-th level. Whereas the quantum-liquid states display the fractional quantum Hall effect, the bubble phases are insulating, and the Hall resistance is thus quantized at integral values of the total filling factor.

\end{abstract}
\maketitle

Two-dimensional electron systems in a strong perpendicular magnetic field display the integer and fractional quantum Hall effects (IQHE and FQHE). \cite{perspectives} At certain values of the ratio $\nu=n_{el}/n_B$ of the electronic $n_{el}$ and the flux densities $n_B=eB/h$, one observes plateaus in the Hall resistance accompanied by a vanishing longitudinal magnetoresistance. The factor $\nu$ determines the filling of the highly degenerate Landau levels (LLs), which are due to the quantization of the electrons' kinetic energy. At integral filling factors $\nu=N$ ($N=2n$ for the lower and $N=2n+1$ for the upper spin branch of the $n$th LL) the non-degenerate ground state, which consists of completely filled levels, is separated from the excited states by a finite gap. By increasing $\nu$ ({\sl e.g.} via lowering $B$) the electrons, which are promoted to higher LLs, become localized due to residual impurities in the sample and therefore do not contribute to the electrical transport. This insulating behavior of electrons gives rise to plateaus of quantized Hall resistance at values $R_{xy}=h/e^2N$ (IQHE). A similar effect (FQHE) arises when the two lowest LLs $n=0$ and $n=1$ are partially filled with $\nubar=\nu-N$. The effect is understood in terms of composite fermions (CFs), which experience a reduced coupling to the magnetic field $(eB)^*=eB/(2ps\pm 1)$, where $s$ is the number of flux pairs carried by each CF and $p$ the number of completely filled CF-LLs. CF localization leads to a quantized Hall resistance at fillings $\nubar=p/(2ps\pm 1)$. However, contrary to the IQHE, the FQHE is entirely caused by electronic interactions. Recent experiments in the $n=1$ LL have revealed an intriguing reentrant IQHE (RIQHE): between the FQHE states at $\nubar=1/5,1/3$ and the even-denominator state at $\nubar=1/2$ with a quantized $R_{xy}=h/e^2(N+\nubar)$, the Hall resistance jumps to values $R_{xy}=h/e^2N$, corresponding to the neighboring plateau of the IQHE. \cite{exp3} The effect is analogous to the RIQHE observed around $\nubar=1/4$ in the $n=2$ LL. \cite{exp2} One-particle localization is unlikely to be at the origin of this insulating phase at such high partial fillings. An alternative origin of the insulating phase is a pinned electron solid such as a charge density wave (CDW) or a Wigner crystal.

In this Rapid Communication we present energy calculations based on a microscopic theory, which show that the RIQHE is due to the formation of a triangular CDW ($M$-electron bubble phase) of electrons in the last partially occupied LL. The energy of the bubble phase is accurately calculated in the Hartree-Fock approximation (HFA) \cite{FKS,moessner,fogler} and is compared to the energy of the quantum liquid, which displays the FQHE, taking into account quasiparticle/quasihole excitations at fillings away from the ``magical'' factors $\nubar_L=1/(2s+1)$ with integral $s$. 

In the high-magnetic-field limit, the Coulomb interaction between the electrons is smaller than the LL separation. Intra-LL excitations, which are possible at fractional filling factors, are therefore more important for the low-energy properties of the system than inter-LL excitations. The latter have been included in a random-phase-approximation calculation and lead to a screened Coulomb interaction in a certain wave vector range. \cite{AG} However, screening has only a minor influence on the physical properties as shown in Ref.\,6, and completely filled LLs may therefore be considered inert. The electronic interactions thus remain as the only energy scale, and one obtains a system of strongly correlated electrons described by the Hamiltonian
\begin{equation}
\label{equ001}
\hat{H}_n=\frac{1}{2}\sum_{\bf q}v(q)\left[F_n(q)\right]^2\bar{\rho}(-{\bf q})\bar{\rho}({\bf q}),
\end{equation}
where $v(q)=2\pi e^2/\epsilon q$ is the two-dimensional Fourier transform of the Coulomb potential. Due to the Zeeman gap, the two branches of the levels $n=1$ and $n=2$ are completely spin-polarized.\cite{exp3,exp2} We therefore consider only interactions between spinless electrons within the $n$th LL described by the density operators
$\langle \rho({\bf q})\rangle_n=F_n(q) \bar{\rho}({\bf q})$,
where $\rho({\bf q})$ is the usual electron density in reciprocal
space. The factors $F_n(q)=L_n(q^2l_B^2/2)\exp(-q^2l_B^2/4)$, with the Laguerre
polynomials $L_n(x)$ and the magnetic length $l_B=\sqrt{\hbar/eB}$, arise from the wave functions of electrons in the $n$th LL and may be absorbed into an effective interaction potential $v_n(q)=v(q)[F_n(q)]^2$. The latter may be interpreted as an interaction potential between electrons whose spatial degrees of freedom are given only in terms of their guiding-center coordinates. The non-commutativity of these coordinates leads to unusual commutation relations for the projected electron density operators, \cite{GMP}
\begin{equation}
\label{equ002}
[\bar{\rho}({\bf q}),\bar{\rho}({\bf k})]=2i\sin\left(\frac{({\bf q}\times{\bf k})_zl_B^2}{2}\right)\bar{\rho}({\bf q}+{\bf k}),
\end{equation}
which together with the Hamiltonian (\ref{equ001}) define the full model. We will set $l_B\equiv 1$ in the following discussion.

The solution of the model in the HFA \cite{FKS,moessner} has predicted a CDW ground state in higher LLs. The cohesive energy is a functional of the order parameter $\Delta(\bq)=\langle\rhobar(\bq)\rangle/n_BA$, where $A$ is the total area of the system,
$$
E_{coh}^{CDW}(n;\nubar)=\frac{n_B}{2\nubar}\sum_{\bq}u_n^{HF}(q)|\Delta(\bq)|^2
$$
with the Hartree-Fock potential $u_n^{HF}(q)=v_n(q)-u_n^F(q)$. The Fock potential is related to $v_n(q)$ by $u_n^F(q)=\sum_{\bp}v_n(p)\exp[i(p_xq_y-p_yq_x)]/n_B$, and one can derive an analytical expression for it,
$$
u_n^F(q)\approx\frac{4e^2}{\epsilon\pi^2n_B q}{\rm Re}\left[K\left(\frac{1-\sqrt{1-4(2n+1)/q^2}}{2}\right)\right]^2,
$$
where $K(x)$ is the complete elliptic integral of the first kind. This formula becomes exact in the large-$n$ limit but scaling arguments have shown that this approximation gives sufficiently accurate results already for $n=1$ and $n=2$.\cite{goerbig2} Even if the HFA fails to describe the quantum liquid phases in the two lowest LLs, it gives correct energy estimates of states with a modulated density such as the electron-solid phases.

In order to describe the bubble phase (triangular CDW) with $M$ electrons per bubble, we use the ansatz $\nubar(\br)=\sum_j \Theta(\sqrt{2M}l_B-|\br-\bf{R}_j|)$ for the local guiding-center filling factor, where the sum is over the lattice vectors of a triangular lattice. \cite{FKS} Fourier transformation of this filling factor yields the order parameter of the bubble phase
$$\Delta_M^B(\bq)=\frac{2\pi\sqrt{2M}}{Aq}J_1(q\sqrt{2M})\sum_j e^{i\bq\cdot{\bf R}_j},$$
where $J_1(x)$ is the first-order Bessel function. The cohesive energy of the $M$-electron bubble phase is thus 
\begin{equation}
\label{equ005}
E_{coh}^{B}(n;M,\nubar)=\frac{n_B\nubar}{M}\sum_l u_n^{HF}({\bf G}_l)\frac{J_1^2(\sqrt{2M}|{\bf G}_l|)}{|{\bf G}_l|^2},
\end{equation}
where ${\bf G}_l\neq 0$ are the reciprocal lattice vectors.

The cohesive energy of the Laughlin states at $\nubar_L=1/(2s+1)$ can be expressed in terms of Haldane's pseudopotentials \cite{haldane1}
$$
E_{coh}^{L}(n;s)=\frac{\nubar_L}{\pi}\sum_{m=0}^{\infty}c_{2m+1}^s V_{2m+1}^{n}
$$
with $V_{2m+1}^{n}=2\pi\sum_{\bq}v_n(q)L_{2m+1}(q^2)\exp(-q^2/2)$. 
The expansion coefficients $c_{2m+1}^s$ may be obtained either from a fit of the 
pair-distribution function of the Laughlin states to Monte-Carlo calculations 
\cite{GMP,laughlin,levesque} or from a certain number of sum rules imposed on 
these coefficients, \cite{GMP,girvin} which result in a system of linear 
equations. \cite{goerbig1} The latter method gives 
results for the energies of the quantum-liquid states which deviate less 
than $1\%$ from Monte-Carlo calculations. \cite{levesque} A systematic 
comparison between the Laughlin states 
and bubble phases in higher LLs has been published by Fogler and Koulakov. 
\cite{fogler} However, their analysis does not permit a comparison between 
the energies of the quantum-liquid and bubble phases away from 
$\nubar_L=1/(2s+1)$, and thus an explanation of the RIQHE is still lacking. 

In order to investigate the energy of the quantum-liquid phase away from 
precisely these filling factors, one has to take into account the excited 
quasiparticles (for $\nubar>\nubar_L$) and quasiholes 
(for $\nubar<\nubar_L$). If the interactions between quasiparticles/quasiholes 
are neglected, one obtains for the cohesive energy of the quantum-liquid 
phases in the $n$-th LL
\begin{equation}
\label{equ007}
E_{coh}^{q-l}(n;s,\nubar)=E_{coh}^{L}(n;s)+[\nubar(2s+1)-1]\Delta^n(s),
\end{equation}
where $\Delta^n(s)$ is the energy of quasiparticles of charge $1/(2s+1)$ (quasiholes of charge $-1/(2s+1)$) in units of the electron charge. $\Delta^n(s)$ can be calculated analytically in the Hamiltonian theory recently proposed by Murthy and Shankar. \cite{MS} The Hamiltonian (\ref{equ001}) is investigated in a CF basis using the ``preferred'' combination 
$$\rhobar^p(\bq)=\rhobar(\bq)-c^2 \chibar(\bq),$$
where $\chibar(\bq)$ is the density operator of a vortex-like excitation of charge $-c^2=-2ps/(2ps+1)$ in units of the electron charge. The choice of the ``preferred'' combination respects the commutation relations (\ref{equ002}) for small wave vectors, whereas the error at larger $q$ is suppressed by the Gaussian in the effective interaction potential. The ground state of this theory is characterized by the expectation value $\langle c_{n,m}^{\dagger}c_{n',m'}\rangle=\delta_{n,n'}\delta_{m,m'}\Theta(p-1-n)$, where $c_{n,m}^{\dagger}$ creates a CF in the $n$-th CF-LL with a CF guiding-center quantum number $m$. The quasiparticle energies are thus given by the expression
$$
\Delta_{qp}^n(s,p)=\langle c_{p,m}\Hhat_n c_{p,m}^{\dagger}\rangle - \langle \Hhat_n \rangle,
$$
and the quasihole energy is
$$
\Delta_{qh}^n(s,p)=\langle c_{p-1,m}^{\dagger}\Hhat_n c_{p-1,m}\rangle - \langle \Hhat_n \rangle,
$$
where one averages over the ground state with the help of Wick contractions. This yields
\begin{eqnarray}
\nonumber
\label{equ008}
\Delta_{qp}^n(s,p)&=&\frac{1}{2}\sum_{\bq}v_n(q)\langle p|\rhobar^p(-\bq)\rhobar^p(\bq)|p \rangle\\
&&-\sum_{\bq}v_n(q)\sum_{n'=0}^{p-1}|\langle p|\rhobar^p(\bq)|n'\rangle|^2
\end{eqnarray}
and 
\begin{eqnarray}
\nonumber
\label{equ009}
\Delta_{qh}^n(s,p)&=&-\frac{1}{2}\sum_{\bq}v_n(q)\langle p-1|\rhobar^p(-\bq)\rhobar^p(\bq)|p-1 \rangle\\
&&+\sum_{\bq}v_n(q)\sum_{n'=0}^{p-1}|\langle p-1|\rhobar^p(\bq)|n'\rangle|^2
\end{eqnarray}
with the matrix elements 
\begin{eqnarray}
\nonumber
\langle p|\rhobar^p(\bq)|n\rangle=\sqrt{\frac{n!}{p!}}\left(\frac{ql_B^{*}c}{\sqrt{2}}\right)^{p-n}e^{-q^2l_B^{*2}c^2/4}\\
\nonumber
\times\left[L_n^{p-n}\left(\frac{q^2l_B^{*2}c^2}{2}\right)-c^{2(1-p+n)}e^{-q^2/2c^2}L_n^{p-n}\left(\frac{q^2l_B^{*2}}{2c^2}\right)\right],
\end{eqnarray}
where $l_B^*=1/\sqrt{1-c^2}$ is the magnetic length for CFs. The expressions (\ref{equ008}) and (\ref{equ009}) are generalizations to an arbitrary LL of Murthy and Shankar's results for $n=0$. \cite{MS} In $n=1$ and $n=2$, one obtains the energies of the quasiparticle excitations for the Laughlin series
\begin{center}
\begin{tabular}{|c||c|c|c|c|}
\hline
$\Delta_{qp}^n(s,p=1)$ & $s=1$ & $s=2$ & $s=3$ & $s=4$\\ \hline \hline
$n=1$ & 0.2267 & 0.1868 & 0.1550 & 0.1316\\ \hline
$n=2$ & 0.1903 & 0.1728 & 0.1543 & 0.1376\\ \hline
\end{tabular}
\end{center}
and the energies of the quasihole excitations
\begin{center}
\begin{tabular}{|c||c|c|c|c|}
\hline
$\Delta_{qh}^n(s,p=1)$ & $s=1$ & $s=2$ & $s=3$ & $s=4$\\ \hline \hline
$n=1$ & -0.07172 & -0.07032 & -0.05887 & -0.04959\\ \hline
$n=2$ & -0.07876 & -0.07853 & -0.06728 & -0.05765\\ \hline
\end{tabular}
\end{center}
in units of $e^2/\epsilon l_B$. 

\begin{figure}
\epsfysize+5.0cm
\epsffile{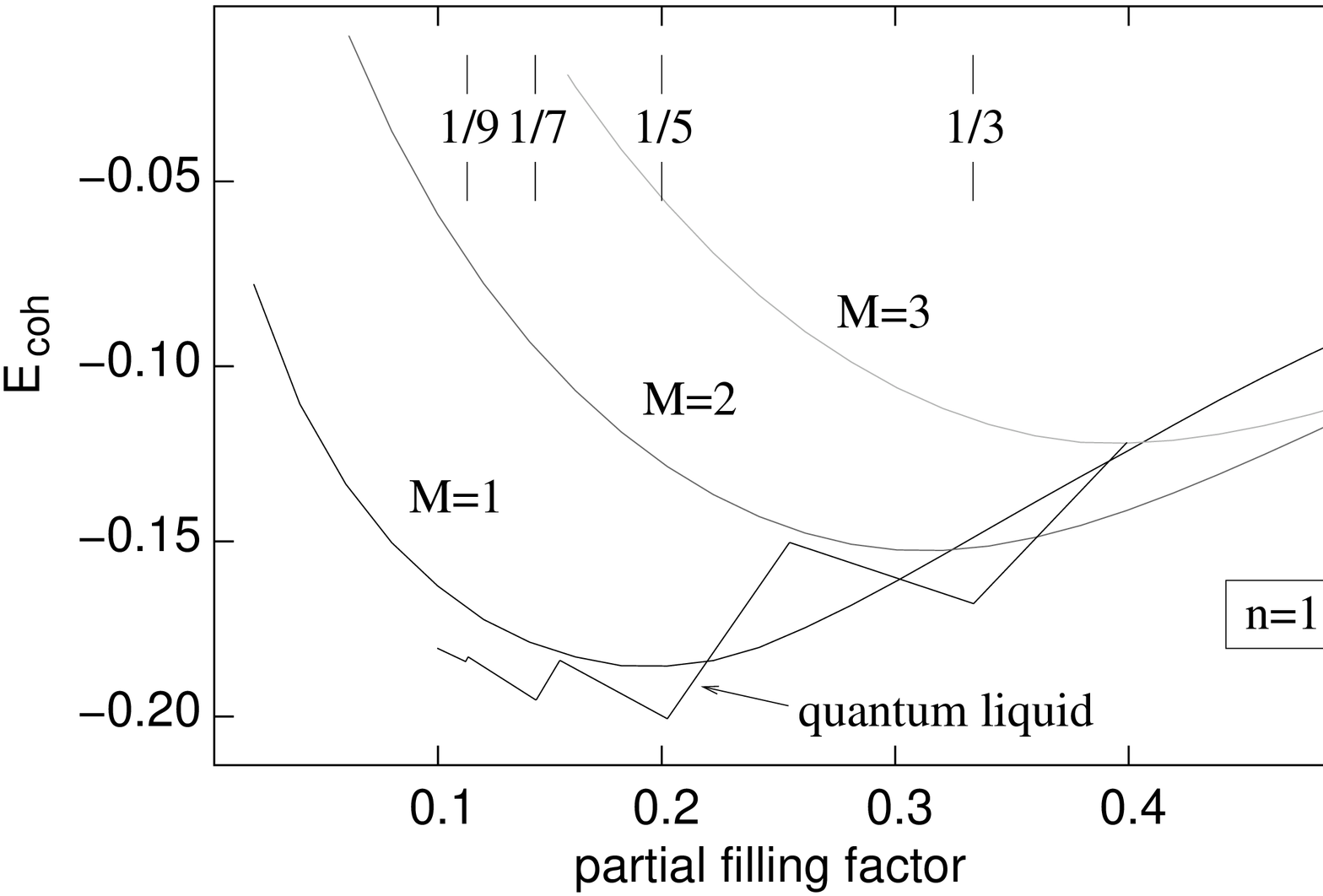}
\epsfysize+5.4cm
\epsffile{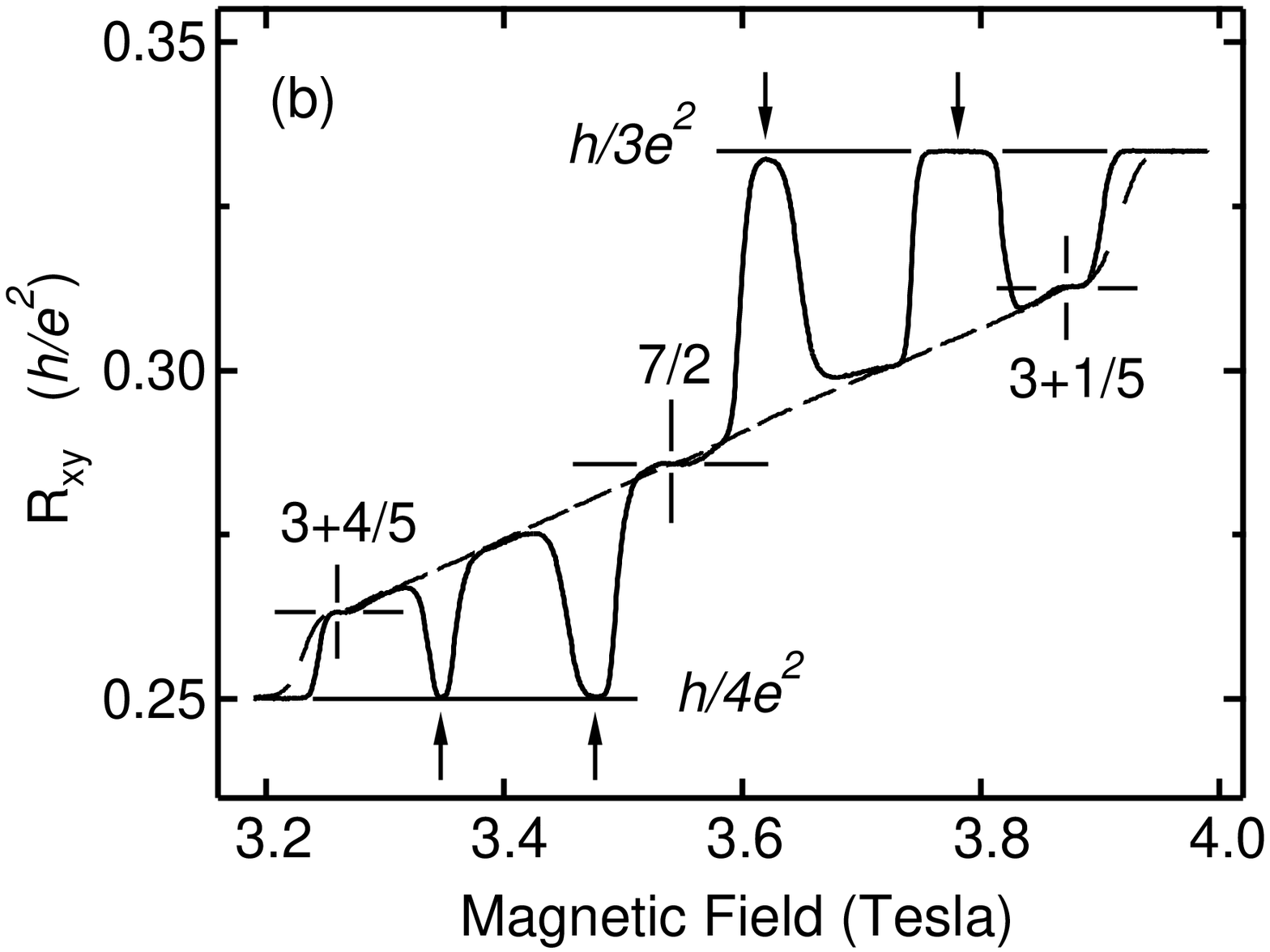}
\caption{(a) Cohesive energies of the $M$-electron 
bubble and quantum-liquid phases for $n=1$ in units of $e^2/\epsilon l_B$.
(b) RIQHE in upper spin branch of $n=1$ measured by Eisenstein {\sl et al.}. 
\cite{exp3}}
\label{fig01}
\end{figure}

\begin{figure}
\epsfysize+5.0cm
\epsffile{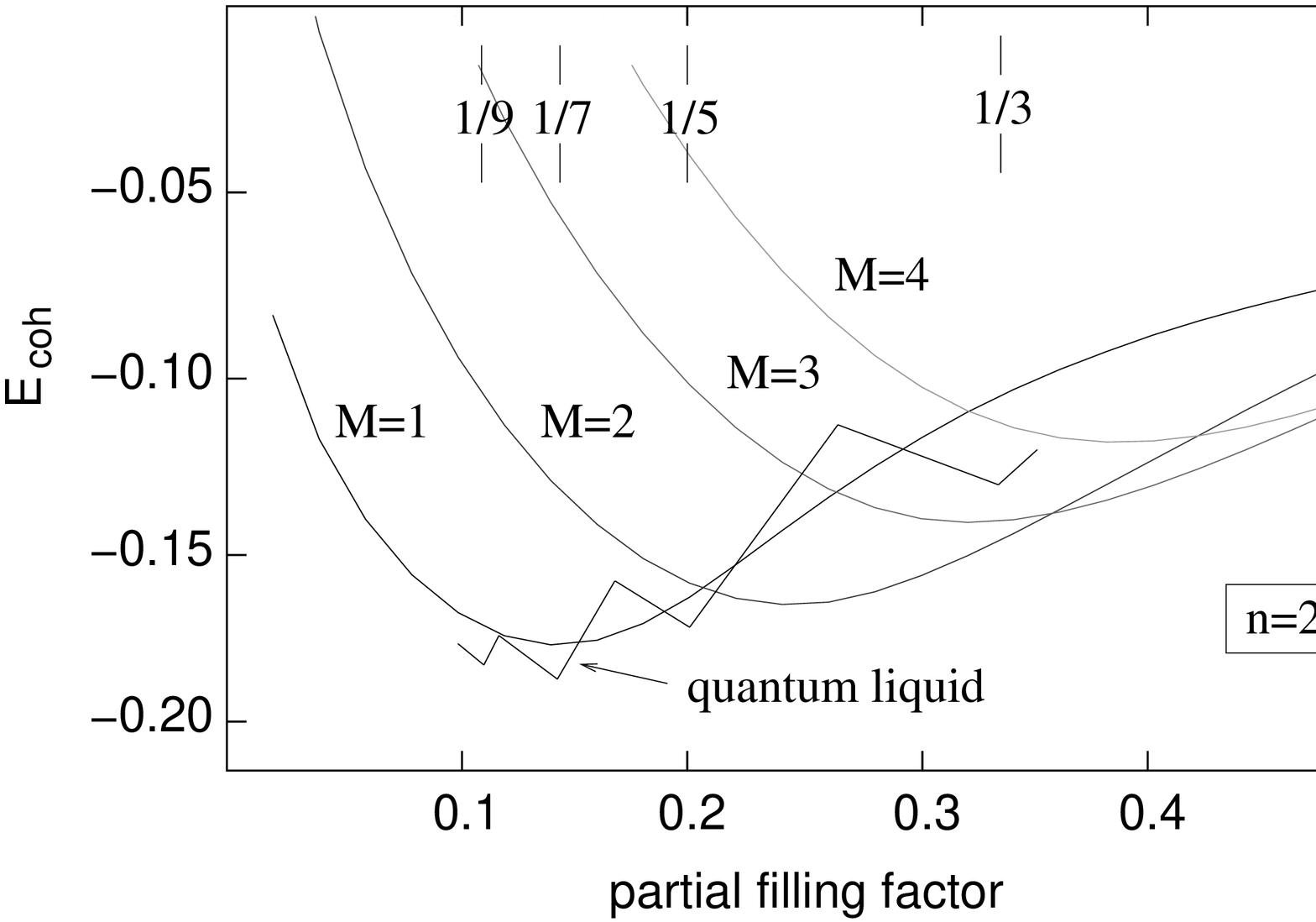}
\epsfysize+5.2cm
\hspace{0.5cm}
\epsffile{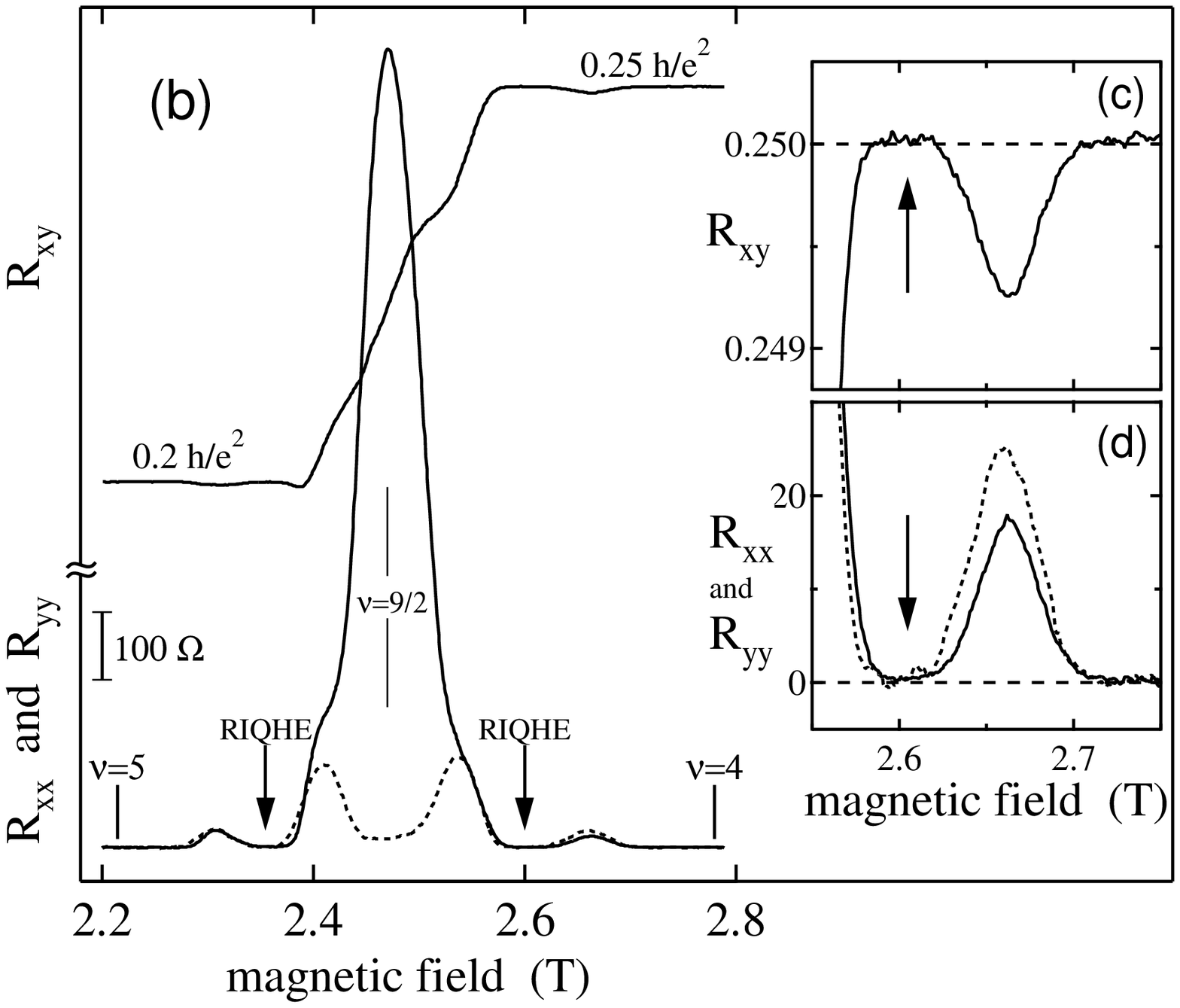}
\caption{(a) Cohesive energy for $n=2$ in units of $e^2/\epsilon l_B$.
(b) RIQHE in lower spin branch of $n=2$ observed by Cooper {\sl et al.}; 
insets are a zoom on Hall (c) and longitudinal (d) resistance around $B=2.65$T.
\cite{exp2}}
\label{fig02}
\end{figure}

A comparison of the cohesive energies of the $M$-electron bubble (\ref{equ005}) 
and the quantum-liquid phases (\ref{equ007}) for $n=1$ and $n=2$ is shown in 
Figs.\,\ref{fig01}a,\,\ref{fig02}a. In the $n=1$ LL the quantum-liquid phases are energetically favorable at $\nubar=1/3\pm 0.03$ and below $\nubar=1/5+0.02$, whereas no liquid phase is found around $\nubar=1/3$ for $n\geq 2$. Note that we have neglected interactions with underlying impurities in the investigation of the bubble phases. Pinning and deformation of the CDW \cite{FL} makes these phases better adapted to follow an underlying electrostatic potential than is an incompressible, homogeneous liquid. The energy gain is more pronounced in the low-density limit, where the elasticity of the CDW is reduced due to a larger lattice constant. This leads to a shift of the curves to smaller $\nubar$. A detailed discussion of this effect will be published elsewhere. \cite{unpubl}
At even lower filling factors, one-particle localization may also destroy the quantum-liquid phase. 

These investigations explain the RIQHE, which was recently 
reported by Eisenstein {\sl et al.} \cite{exp3} for $n=1$ (Fig.\,\ref{fig01}b) 
and which had been observed before by Cooper {\sl et al.} \cite{exp2}
in $n=2$ around $\nubar=1/4$ (Fig.\,\ref{fig02}b). The insulating behavior of electrons in the last LL, which is responsible for the RIQHE, may be caused by simple one-particle localization or by collective effects such as crystallization into an electron solid. Electrical transport in an electron-solid phase arises either by a collective sliding, which is suppressed by pinning due to residual impurities, or by the propagation of crystal dislocations, which is reduced in the limit $T\rightarrow 0$. The observation of the RIQHE at rather high partial fillings, comparable to those at which the FQHE is found, indicates the relevance of Coulomb interactions. This favors an explanation in terms of the electron solid, {\sl i.e.} $M$-electron bubble phases, as being responsible for the insulating behavior. This picture is supported by the energy investigations presented above. 

In the $n=1$ LL the FQHE, as a property of the quantum-liquid phases, 
is observable at $\nu=2+1/3$ and $2+1/5$ (lower spin branch) and at 
$\nu=3+1/3$ and $3+1/5$ (upper spin branch, {\sl c.f.} Fig.\,\ref{fig01}b), 
whereas between these FQHE states the energetically favored $1$-electron 
bubble phase gives rise to the observed RIQHE. At $\nubar\geq 1/3+0.03$, 
one observes again the RIQHE caused by a $2$-electron bubble phase 
(Fig.\,\ref{fig01}a) in perfect agreement with the experiment \cite{exp3} 
and recent numerical density-matrix-renormalization-group studies. \cite{shibata} 

In the $n=2$ LL, the RIQHE around $\nubar=1/4$ is due to the existence of 
a $2$-electron bubble phase and $3$-electron bubbles for $\nubar\geq 0.35$. 
The small maxima in the longitudinal magnetoresistance, which are observed 
at $B=2.32$T and $B=2.65$T corresponding to filling factors $\nu=4+4/5$ and 
$\nu=4+1/5$, \cite{exp2} may indicate an incipient quantum melting of the 
bubble phase. Such a quantum melting is expected from the energy 
investigations, which indicate a quantum-liquid ground state and thus a 
possible observation of the FQHE at $\nubar=1/5$ or $1/7$ in $n=2$ 
(Fig.\,\ref{fig02}a). \cite{goerbig2} 

In higher LLs $n\geq 3$ (results not shown), the bubble phases are of 
lower energy than the FQHE state at $\nubar=1/5$.  One observes a 
general shift of the quantum-liquid phases to lower partial filling 
factors with increasing LL index $n$ in agreement with results from 
scaling investigations. \cite{moessner,goerbig2}

In conclusion, we have presented a mechanism for the appearance of 
the RIQHE in the $n=1$ and $n=2$ LLs based on the comparison between 
the energies of the different $M$-electron bubble and 
quantum-liquid phases. Whereas the energies of the bubble phases are 
calculated in the HFA, which gives reliable results for electronic 
states with a modulated density, the quantum-liquid phases are 
characterized by quantum correlations beyond the mean-field level. 
The energies of the Laughlin liquid are calculated to great accuracy 
in an approach based on physical sum rules, \cite{goerbig1} and the 
energies of quasiparticles/quasiholes, which are excited at fillings 
away from the ``magical'' factors $\nubar_L=1/(2s+1)$ are obtained 
in the framework of the Hamiltonian theory. \cite{MS} In a certain 
range of the partial filling factor, one finds an alternating 
sequence of $M$-electron-bubble and quantum-liquid ground states 
with first-order quantum phase transitions between them. Whereas 
the quantum-liquid states display the FQHE with a quantized Hall 
resistance $R_{xy}=h/e^2\nu$, $\nu=N+\nubar_L$, 
the bubble phases are insulating, and the electrical transport 
is thus entirely due to the completely filled lower LLs. This 
gives rise to the RIQHE with a quantized Hall resistance 
$R_{xy}=h/e^2N$ independent of the partial filling factor, as is 
observed experimentally. \cite{exp2,exp3}

We acknowledge fruitful discussions with K.\ Borejsza, J.\ P.\ Eisenstein, R.\ Moessner, and V.\ Pasquier. This work was supported by the Swiss National Foundation for Scientific Research under grant No.~620-62868.00.


\begin{thebibliography}{99}

\bibitem{perspectives}For a review, see {\sl Perspectives in Quantum Hall Effects}, edited by S.\ Das\ Sarma and A.\ Pinczuk (John Wiley, New York, 1997).

\bibitem{exp3}J.\ P.\ Eisenstein, K.\ B.\ Cooper, L.\ N.\ Pfeiffer, and K.\ W.\ West, Phys.\ Rev.\ Lett.\ {\bf 88}, 076801 (2002).

\bibitem{exp2}K.\ B.\ Cooper, M.\ P.\ Lilly, J.\ P.\ Eisenstein, L.\ N.\ Pfeiffer, and K.\ W.\ West, Phys.\ Rev.\ B\ {\bf 60}, 11285 (1999).

\bibitem{FKS}A.\ A.\ Koulakov, M.\ M.\ Fogler, and B.\ I.\ Shklovskii, Phys.\ Rev.\ Lett.\ {\bf 76}, 499 (1996);
M.\ M.\ Fogler, A.\ A.\ Koulakov, and B.\ I.\ Shklovskii, Phys.\ Rev.\ B\ {\bf 54}, 1853 (1996).

\bibitem{moessner}R.\ Moessner and J.\ T.\ Chalker, Phys.\ Rev.\ B\ {\bf 54}, 5006 (1996).

\bibitem{fogler}M.\ M.\ Fogler and  A.\ A.\ Koulakov, Phys.\ Rev.\ B\ {\bf 55}, 9326 (1997).

\bibitem{AG}I.\ L.\ Aleiner and L.\ I.\ Glazman, Phys.\ Rev.\ B\ {\bf 52}, 11296 (1995).

\bibitem{GMP}S.\ M.\ Girvin, A.\ H.\ MacDonald, and P.\ M.\ Platzman, Phys.\ Rev.\ B\ {\bf 33}, 2481 (1986).

\bibitem{goerbig2}M.\ O.\ Goerbig and C.\ Morais\ Smith, Europhys.\ Lett.\ {\bf 63}, 736 (2003).

\bibitem{haldane1}F.\ D.\ Haldane, Phys.\ Rev.\ Lett.\ {\bf 51}, 605 (1983).

\bibitem{laughlin}R.\ B.\ Laughlin, Phys.\ Rev.\ Lett.\ {\bf 50}, 1395 (1983).

\bibitem{levesque}D.\ Levesque, J.\ J. Weis, and A.\ H.\ MacDonald, Phys.\ Rev.\ B\ {\bf 30}, 1056 (1984).

\bibitem{girvin}S.\ M.\ Girvin, Phys.\ Rev.\ B\ {\bf 30}, 558 (1984).

\bibitem{goerbig1}M.\ O.\ Goerbig and C.\ Morais\ Smith, Phys.\ Rev.\ B\ {\bf 66}, 241101 (2002).

\bibitem{MS}G.\ Murthy and R.\ Shankar, Rev.\ Mod.\ Phys.\ {\bf 75}, 1101 (2003); R.\ Shankar, Phys.\ Rev.\ B\ {\bf 63}, 085322 (2001).

\bibitem{FL} H.\ Fukuyama and P.\ A.\ Lee, Phys.\ Rev.\ B\ {\bf 17}, 535 (1977).

\bibitem{unpubl}M.\ O.\ Goerbig, P.\ Lederer, and C.\ Morais\ Smith, {\sl unpublished}.

\bibitem{shibata}N.\ Shibata and D.\ Yoshioka, J.\ Phys.\ Soc.\ Jpn.\ {\bf 72}, 664 (2003).


\end{thebibliography}
\end{document}